\documentclass[pra,10pt,twocolumn,superscriptaddress,floatfix,showpacs]{revtex4-1}

\usepackage{subfig}
\usepackage{graphicx} 
\usepackage{epstopdf}
\usepackage[utf8]{inputenc}
\usepackage[T1]{fontenc}     
\usepackage[british]{babel}  
\usepackage[sc,osf]{mathpazo}
\usepackage[scaled=0.86]{berasans}  
\usepackage[colorlinks=true, citecolor=blue, urlcolor=blue]{hyperref}  
\usepackage[babel]{microtype}  
\usepackage{amsmath,amssymb,amsthm,bm,amsfonts,mathrsfs,bbm} 

\usepackage{xspace}  
\usepackage{pgfplots}
\usepackage{xcolor,colortbl}
\def\ba{\begin{equation}}
\def\ea{\end{equation}}
\def\bea{\begin{eqnarray}}
\def\eea{\end{eqnarray}}
\def\ben{\begin{equation*}}
\def\een{\end{equation*}}
\def\bean{\begin{eqnarray*}}
\def\eean{\end{eqnarray*}}
\def\bma{\begin{mathletters}}
\def\ema{\end{mathletters}}
\def\bi{\begin{itemize}}
\def\ei{\end{itemize}}

\newcommand{\be}{\begin{equation}}
\newcommand{\ee}{\end{equation}}

\newcommand{\kommentar}[1]{}

\newcommand{\forget}[1]{}

\begin{document}

\title{Restricted Distribution of Quantum Correlations in Bilocal Network}

\author{Kaushiki Mukherjee}
\email{kaushiki_mukherjee@rediffmail.com}
\affiliation{Department of Mathematics, Government Girls' General Degree College, Ekbalpore, Kolkata-700023, India.}

\author{Biswajit Paul}
\email{biswajitpaul4@gmail.com}
\affiliation{Department of Mathematics, South Malda College, Malda, West Bengal, India}

\author{Debasis  Sarkar}
\email{dsarkar1x@gmail.com, dsappmath@caluniv.ac.in}
\affiliation{Department of Applied Mathematics, University of Calcutta, 92, A.P.C. Road, Kolkata-700009, India.}


\begin{abstract}
Analyzing shareability of correlations arising in any physical theory may be considered as a fruitful technique of
studying the theory. Our present topic of discussion involves an analogous approach of studying quantum theory. For
our purpose, we have deviated from the usual procedure of assessing monogamous nature of quantum correlations in
standard Bell-CHSH scenario. We have considered correlations arising in a quantum network involving independent sources.
Precisely speaking, we have analyzed monogamy of nonbilocal correlations by deriving a relation restricting marginals. Interestingly, restrictions constraining distribution of nonbilocal correlations remain same irrespective of whether inputs of the nodal observers are kept fixed(in different bilocal networks) while studying nonbilocal nature of marginal correlations.
\end{abstract}

\maketitle

	
\section{Introduction}
Entanglement and nonlocality, the two most intrinsic features of quantum theory, play ubiquitous role in analyzing
departure of the theory from the classical world. While the former is a property of quantum states\cite{review}, the
 latter mainly characterizes nature of correlations arising due to measurements on quantum
 systems\cite{Bel},\cite{Bel2},\cite{Brunreview}. Considered to be two inequivalent resources in general, both of these
  features form the basis of various information processing tasks such as device independent entanglement witnesses\cite{Bancal},
  Quantum Key Distribution(QKD)\cite{Acin,key1,Mayer,key2}, Bayesian game theoretic applications \cite{game}, private randomness
   generation\cite{Pironio,Colbeck}, etc, which cannot be performed by any classical resource. One of the inherent
 features responsible for strengthening efficiency of quantum resources over classical ones is the existence of restrictions over shareability of quantum particles or quantum correlations in multiparty
     scenario\cite{monqb,monc,monqd,monqt,monqp,monqk,monqo,monj,mons,mony,mong,monk}.\\
Research activities conducted so far clearly point out the existence of limitations over shareability of both quantum
nonlocality\cite{monqb,monqd} and entanglement\cite{monc,monj,mons,mony,mong,monk}. Such sort of limitations are
frequently referred to as monogamy of nonlocality and entanglement respectively. Precisely speaking, let a tripartite
state be shared between three parties, say, Alice, Bob and Charlie. If Alice's qubit is maximally entangled with that
of Bob, then neither the state shared between Alice and Charlie nor that between Bob and Charlie is entangled. Now
consider the tripartite correlations generated due to measurements on a quantum system shared between Alice, Bob and
 Charlie. If the marginal correlations shared between any two parties, say Alice and Bob violate Bell-CHSH
 inequality\cite{Cl} maximally then neither marginal shared between Alice and Charlie nor that shared between Bob and
 Charlie can show Bell-CHSH violation. However no such restriction exists over shareability of classical correlations.
 Over years, different trade-off relations have been designed to capture monogamous nature of not only quantum
 correlations but also of correlations abiding by no signaling principle\cite{ns}. Our present topic of discussion is
  contributory in this direction. To be precise, we have explored shareability of correlations characterizing quantum
  \textit{bilocal network}.\\
Over past few years, there has been a trend of studying quantum networks involving independent sources\cite{BRA,BRAN}.
 Networks involving two independent sources are referred to as `bilocal' network. It was first introduced in \cite{BRA}.
  Since then study of quantum networks characterized with source independence has been subject matter of thorough
  investigations\cite{Tav,raf,km,den,kmb,kmbs,gis1,raf1} due to multi-faceted utility of the source independence
  assumption both from theoretical and experimental perspectives. For instance, it can be exploited to lower down
  restrictions for detecting quantumness(nonclassical feature) in a network via some notions of quantum nonlocality
  (different from standard Bell nonlocality)\cite{kmb,kmbs}. Besides, it is found to be important to study detection   loophole in some local models\cite{Gisin,Greenberger}. From experimental perspectives, source independent networks
   form basis of various experiments related to quantum information and communication such as various device independent
    quantum information processing tasks\cite{Mayer,key2,Bancal,game}, some communication networks dealing with entanglement
     percolation \cite{network1}, \textit{quantum repeaters} \cite{repeat} and \textit{quantum memories} \cite{memory}, etc.
     Owing to the significance of these networks, study of correlations generated in such networks have gained immense importance. In this context, one obvious direction of investigation evolves around manifesting shareability of correlations in such networks. Our discussions will channelize in that direction.\\
To the best of authors' knowledge, research activities on monogamy of quantum correlations, conducted so far, basically
 considers standard Bell scenario. Here we have shifted from that usual scenario thereby exploring the same for quantum
 correlations in a network scenario. Precisely speaking, we have considered quantum network involving two independent sources
  with an urge to investigate whether non classical feature of quantum correlations generated in such networks exhibit monogamy
  or not. We have obtained affirmative answer to this query. It may be noted that for studying monogamy in standard Bell-CHSH scenario,
   it is assumed that the nodal party(for instance Alice in the example discussed before) has fixed measurement settings.
   For instance, to analyze Bell violation by each of two sets of bipartite correlations: $P(a,b|x,y)$, shared between Alice, Bob
    and $P(a,c|x,z)$, shared between Alice, Charlie($a,b,c$ and $x,y,z$ denoting binary outputs and inputs of Alice, Bob
     and Charlie respectively), Alice's measurement settings are assumed to be fixed. Recently, in \cite{fei1}, a trade-off relation
     has been suggested giving restriction over upper bound of violation of Bell-CHSH violation by all the possible bipartite marginals
     where the measurement settings of nodal party was not assumed to be fixed. Here we have started deriving a monogamy relation for
     nonbilocal correlations. Then we have relaxed the assumption of fixed setting by nodal party, thereby designing a trade-off relation
      restricting the nonbilocal nature of the marginal correlations. Interestingly, nature of restrictions to exhibit nonbilocality by
      the marginals remain invariant irrespective of the assumption of fixed measurement settings of nodal party. \\
Rest of the article is organized as follows. First we discuss some ideas motivating our work in Sec.\ref{mot}. Next in
 Sec.\ref{pre}, we give a brief review of the bilocal network scenario and some results related to that scenario which
 in turn will facilitate our further discussions. In Sec.\ref{mono}, we first sketch the network scenario in details.
 Then we derive the monogamy relation in Sec.\ref{mono}  followed by a trade-off relation in Sec.\ref{trade} restricting
  the correlations generated therein. Some practical implications of our findings have been discussed in Sec.\ref{prac}.
  Finally we have concluded in Sec.\ref{conc} discussing possible future directions of research activities.
\section{Motivation}\label{mot}
As we have already pointed out before that in recent times, study of quantum networks(with independent sources) has
gained paramount interest. So detailed analysis of various aspects of correlations generated in such networks enriches
quantum theory. Again monogamy is one of the most important nonclassical feature of quantum theory. So assessment of monogamous
nature(if any) of correlations arising in such quantum networks is crucial for developing a better insight in related fundamental
issues.\\
From practical view point, existence of restrictions over shareability of quantum correlations is utilized to design quantum
secret sharing protocol secure against eavesdropping\cite{cryp1,cryp2,cryp3}. To be specific, it is this nonclassical feature
of quantum correlations that plays a vital role to provide security against external attack better than any classical protocol.
So if monogamous nature of correlations arising in quantum networks involving independent sources can be guaranteed then that will
be definitely helpful for security analysis in secret sharing protocols involving such networks. So possible issues related with
restricted shareability of quantum correlations in network scenario deserves detailed investigations. This basically motivates
our current topic.
\section{Preliminaries}\label{pre}
\subsection{Bilocal Scenario}
Bilocal network as designed in \cite{BRA,BRAN} is shown in Fig.1. The network involves three parties Alice($A$), Bob($B$)
and Charlie($C$) and two sources $\textbf{S}_1$ and $S_2.$ All the parties and sources are arranged in a linear fashion.
A source is shared between any pair of adjacent parties. Sources $\textbf{S}_1$ and $\textbf{S}_2$ are independent to each
other($\textit{bilocal assumption}$). A physical system represented by variables $\lambda_1$ and $\lambda_2$ is send
by $\textbf{S}_1$ and $\textbf{S}_2$ respectively. Bob receives two particles(one form each of $\textbf{S}_1$ and $\textbf{S}_2$). Independence of $\lambda_1$
and $\lambda_2$ is ensured by that of $\textbf{S}_1$ and $\textbf{S}_2$. Each of Alice, Bob and Charlie can perform dichotomic measurements
 on their systems. The binary inputs are denoted by $x,\,y,\,z$ for Alice, Bob and Charlie and their outputs are labeled
 as $a,\,b,\,c$ respectively. In particular, Bob performs measurement on the joint state of the two systems that he receives
  from $\textbf{S}_1$ and $\textbf{S}_2$. The correlations obtained in the network are local if they take the form:
$P(a, b, c|x, y, z)=\iint d\lambda_{1} d\lambda_{2} \rho(\lambda_1,\lambda_2)$
\begin{equation}\label{p11}
P(a|x, \lambda_1)P(b|y, \lambda_1, \lambda_2)P(c|z, \lambda_2)
\end{equation}
Tripartite correlations $P(a, b, c|x, y, z)$ are bilocal if they can be decomposed in above form(Eq.(\ref{p11}))
 together with the restriction:
\begin{equation}\label{p2}
    \rho(\lambda_1,\lambda_2)=\rho_1(\lambda_1)\rho_2(\lambda_2)
\end{equation}
imposed on the probability distributions of the hidden variables $\lambda_1, \lambda_2$. Eq.(\ref{p2}) refers to
the \textit{bilocal constraint}.
\begin{figure}[htb]
\includegraphics[width=3in]{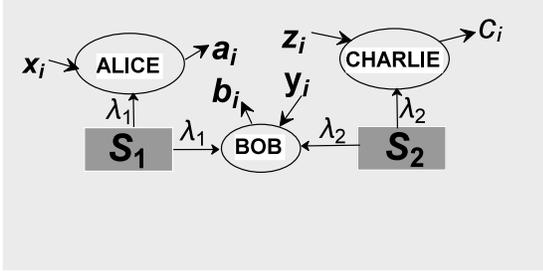}\\
\caption{\emph{Schematic diagram of a bilocal network\cite{BRAN,BRA}. }}
\end{figure}
Tripartite correlations of the form (Eq.(\ref{p11})) and (Eq.(\ref{p2})) are bilocal if they satisfy the inequality\cite{BRAN}:
\begin{equation}\label{A3}
 \sqrt{|I|} + \sqrt{|J|}\leq1
\end{equation}
$\textmd{where}\,\,I$=$\frac{1}{4}\sum \limits_{x, z=0,1}\langle A_x B_0 C_z\rangle,\,J$=
$\frac{1}{4}\sum \limits_{x, z=0, 1}(-1)^{x+z}\langle A_x B_1 C_z\rangle\,$ $\textmd{and}\,\langle A_x B_y C_z\rangle$=
$\sum\limits_{a, b, c}(-1)^{a+b+c}P(a, b, c|x,y, z).$ $A_x$, $B_y$ and $C_z$ stand for the observables corresponding
to binary inputs $x,\,y,\,z$ of Alice, Bob and Charlie respectively. $a,b,c\in\{0,1\}$ denote the corresponding outputs.
 Denoting $ \sqrt{|I|} + \sqrt{|J|}$ as $\mathbf{B},$ Eq.(\ref{A3}) becomes:
\begin{equation}\label{A3l}
\mathbf{B}\leq1
\end{equation}
Clearly violation of Eq.(\ref{A3l}) acts as a sufficient criterion for detecting nonbilocality of corresponding
correlations.
 In \cite{raf1}, referring $\mathbf{B}$ as \textit{bilocality parameter}, an upper bound of quantum violation of
  the bilocal inequality(Eq.(\ref{A3l})) has been derived under the assumption that Bob performs separable measurements
  on the joint state of its two particles. We next briefly review scenario considered in \cite{raf1} along with some of
  the related findings which will be used later in course of our work.
\subsection{Bilocal Quantum Network}
Let each of $\textbf{S}_1$ and $\textbf{S}_2$ sends a quantum state. Let $\textbf{S}_1$ sends $\rho_{AB}$ to Alice and Bob whereas $\textbf{S}_2$ sends
 $\rho_{BC}$ to Bob and Charlie. In general any bipartite state density matrix representing a quantum state($\rho$, say) can be defined as:
\begin{equation}\label{st4}
\rho=\frac{1}{2^2}\sum_{i_1,i_2=0}^{3}T_{i_1i_2}\sigma^1_{i_1}\bigotimes\sigma^2_{i_2}
\end{equation}
with $\sigma^k_0,$ denoting the identity operator in the Hilbert space of $k^{th}$ qubit and $\sigma^k_{i_k},$ denote
 the Pauli operators along three mutually perpendicular directions, $i_k=1,2,3$.
The entries of the correlation matrix $T_{\rho}$ of $\rho$ denoted by $T_{i_1i_2},$ are real and given by:
 \begin{equation}\label{st4i}
 T_{i_1i_2}=\textmd{Tr}[\rho\sigma^1_{i_1}\bigotimes\sigma^2_{i_2}],\,i_1,i_2\in\{1,2,3\}.
 \end{equation}
Each of the three parties performs projective measurements in arbitrary directions. Alice and Charlie performs:
 $\vec{\alpha_i}.\vec{\sigma}$ and $\vec{\gamma_i}.\vec{\sigma}(i=0,1)$ respectively. Here $\vec{\sigma}$$=$$(\sigma_1,\sigma_2,\sigma_3)$.
  Bob performs separable measurements on its two qubits\cite{raf1}: $\vec{\beta_i^A}.\vec{\sigma}\otimes\vec{\beta_i^C}.\vec{\sigma}.$
  Under these settings, bilocality parameter $\mathbf{B}$ takes the form\cite{raf1}:
\begin{equation}\label{bm1}
\mathbf{B}=\frac{1}{2}\small{\sum}_{i=0}^1\sqrt{|(\vec{\alpha_0}+
(-1)^i\vec{\alpha_1}).T_{AB}\vec{\beta_i^A}||\vec{\beta_i^C}.T_{BC}(\vec{\gamma_0}+
(-1)^i\vec{\gamma_1})|},
\end{equation}
where $T_{AB}$ and $T_{BC}$ denote correlation tensor of state $\rho_{AB}$ and $\rho_{BC}$ respectively.
The upper bound of violation of the bilocal inequality(Eq.(\ref{A3l})) is given by $\mathbf{B}_{Max}$\cite{raf1}:
\begin{equation}\label{bm2}
 \mathbf{B}_{Max}=\sqrt{\small{\sum_{i=1}^2}\sqrt{\omega_i^A*\omega_i^C}},\,\, \,\omega_1^{A(C)}>\omega_2^{A(C)},
\end{equation}
with $\omega_1^A$ and $\omega_2^A(\omega_1^C\, \textmd{and}\,\omega_2^C)$ are the larger two eigenvalues
of $T_{AB}^TT_{AB}(T_{BC}^TT_{BC}).$\\
After discussing the mathematical pre-requisites, we proceed to present our findings.
\section{Nonbilocal Monogamy}\label{mono}
For exploring restriction(if any) over shareability of nonbilocal correlations, first we define the network
 scenario(Fig.2).\\
\begin{figure}[htb]
\includegraphics[width=3in]{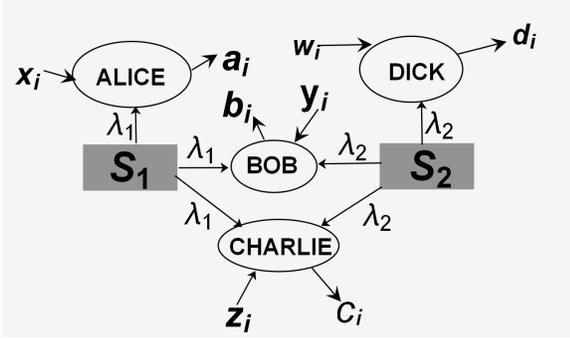}\\
\caption{\emph{Schematic diagram of network $\mathcal{N}$. }}
\end{figure}
Consider a network($\mathcal{N}$) involving four parties Alice, Bob, Charlie, Dick and two independent sources
$\textbf{S}_1$ and $\textbf{S}_2$. Each of the two sources generates a tripartite quantum state. Let $\textbf{S}_1$ generates $\rho_{ABC}$,
sending a qubit to each of Alice, Bob and Charlie. Analogously, $\textbf{S}_2$ generates $\rho_{BCD}$, sending a qubit to
each of Bob, Charlie and Dick. So each of Bob and Charlie receives two qubits whereas remaining two parties receives one qubit each. Let Bob and Charlie be referred to as \textit{intermediate parties} whereas Alice and Dick be referred to as \textit{extreme parties}. The extreme parties perform arbitrary projective measurements locally on their
  qubits and each of two intermediate parties perform separable measurements on joint state of its two qubits.
  Inputs of Alice, Bob, Charlie and Dick are labeled as $x,y,z,w\in\{0,1\}$ and outputs as $a,b,c,d\in\{0,1\}$
  respectively. Four partite correlation terms $P(a,b,c,d|x,y,z,w),$ arising due to measurements by the parties
  on their respective qubits characterize the network($\mathcal{N}$). Let $W_B$ and $W_C$ denote the set of tripartite  marginals $P(a,b,d|x,y,w)$ and $P(a,c,d|x,z,w)$ respectively. Now $W_B$ can be interpreted as the set of tripartite
  correlations arising due to binary measurements by each of three parties Alice, Bob and Dick in a network involving
  two independent sources $\textbf{S}_1$ and $\textbf{S}_2.$ Hence, correlations from $W_B$ characterize the bilocal network
  ($\mathcal{N}_B$, say) involving Alice, Bob and Dick. Analogously $W_C$ characterize bilocal network
  ($\mathcal{N}_C$, say)
involving parties Alice, Charlie and Dick. Each of the two bilocal networks $\mathcal{N}_B$ and $\mathcal{N}_C$
may be referred to as a \textit{reduced network} obtained from the original bilocal network $\mathcal{N}.$
Clearly extreme parties of $\mathcal{N}$ are common in both the reduced networks($\mathcal{N}_B,$ $\mathcal{N}_C$)
and may be referred to as the \textit{nodal parties}.\\
Now we put forward the monogamy relation restricting nonbilocality of the tripartite marginals $P(a,b,d|x,y,w)$ and $P(a,c,d|x,z,w).$\\
\textit{Theorem.1}: If $\mathbf{B}_{Max}^B$ and $\mathbf{B}_{Max}^C$ denote the upper bound of violations of bilocal
inequality(Eq.(\ref{A3l})) by correlations $P(a,b,d|x,y,w)$ and $P(a,c,d|x,z,w)$ respectively, then,
\begin{equation}\label{bm3}
    (\mathbf{B}_{Max}^B)^2+(\mathbf{B}_{Max}^C)^2\leq 2.
\end{equation}
\textit{Proof}: See appendix.\\
Compared to a single nodal(common) party in standard Bell scenario, here measurement settings of both of the nodal parties (Alice and Dick) are kept fixed in order to sketch the monogamy relation(Eq.(\ref{bm3})). Alice and Dick's fixed measurement settings mainly refer to the fact each of their measurement settings remain unchanged in both the reduced networks $\mathcal{N}_B,$ $\mathcal{N}_C.$ To be precise, if Alice(Dick) performs measurement $M_A(M_D)$ in network $\mathcal{N}_B$, then in network $\mathcal{N}_C$ also measurement setting of Alice(Dick) is $M_A(M_D).$ Also note that the monogamy relations are sketched here under the assumption that Bob and Charlie perform separable measurements.  \\
\textit{Tightness of the constraint:} By tightness of the monogamy relation given by Eq.(\ref{bm3}) we interpret the existence of quantum correlations reaching the upper bound $2.$ For a particular instance, let each of the two sources $\textbf{S}_1$ and $\textbf{S}_2$ generates identical copy of a $W$ state\cite{ACN}:
\begin{equation}\label{state}
    |\Psi\rangle=\cos\mu_0|001\rangle +\sin\mu_1\sin\mu_0 |010\rangle + \sin\mu_0\cos\mu_1|100\rangle
\end{equation}
where $\mu_i(i=0,1)$$\in$$[0,\frac{\pi}{2}].$ Let, for $\mu_0=\frac{\pi}{2}$ identical copies of the corresponding state($|\Psi\rangle\langle\Psi|$) are used in the networks($\mathcal{N}_B,$ $\mathcal{N}_C$). Maximizing over all possible separable measurement settings of Bob, Charlie and all projective measurement settings of Alice and Dick(as already argued, each of Alice and Dick performs same measurement in both the networks $\mathcal{N}_B,$ $\mathcal{N}_C$), we get $(\mathbf{B}_{Max}^B)^2+
 (\mathbf{B}_{Max}^C)^2$$=$$\sqrt{(\textmd{Max}[0,\cos^2(2\mu_1)])^2}+\sqrt{(\textmd{Max}[1,\sin^2(2\mu_1)])^2}+
 \sqrt{(\textmd{Min}[0,\cos^2(2\mu_1)])^2}+\sqrt{(\textmd{Min}[1,\sin^2(2\mu_1)])^2}.$ Clearly on simplification, $(\mathbf{B}_{Max}^B)^2+
 (\mathbf{B}_{Max}^C)^2$$=$$2.$ \\
Before discussing any further observation, we first put forward a lemma. \\
\textit{Lemma}: Maximal violation of the bilocal inequality(Eq.\ref{A3l}) is $\sqrt{2}$,
maximum being taken over all possible quantum states.\\
\textit{Proof}: From upper bound of violation of bilocal inequality(Eq.(\ref{A3l})) given by Eq.(\ref{bm2}),
\begin{equation}\label{bm4}
    \mathbf{B}_{Max}^2=\small{\sum_{i=1}^2}\sqrt{\omega_i^A*\omega_i^C}.
\end{equation}
By Cauchy-Schwarz's inequality, we get
\begin{equation}\label{bm5}
\mathbf{B}_{Max}^2\leq \sqrt{\omega_1^A+\omega_2^A}\sqrt{\omega_1^C+\omega_2^C}
\end{equation}
Now $\sqrt{\omega_1^A+\omega_2^A}=M(\rho_{AB})$ and likewise $\sqrt{\omega_1^C+\omega_2^C}=M(\rho_{BC})$ where $2M(\rho)$ denote maximal violation of Bell-CHSH inequality by a quantum state $\rho$\cite{hord1}.
Again maximal possible quantum violation of Bell-CHSH is given by $2\sqrt{2},$ referred
to as Tsirelson's bound\cite{Tir}. Hence maximal possible quantum violation of the bilocal
inequality(Eq.(\ref{A3l})) turns out to be $\sqrt{2}.$ $\blacksquare$\\
The monogamy relation(Eq.(\ref{bm3})) puts restrictions over distribution of nonbilocal correlations among
the two networks $\mathcal{N}_B$ and $\mathcal{N}_C.$ To be precise, maximal violation of bilocal
inequality(Eq.(\ref{A3l})), being $\sqrt{2},$ both of $(B_{Max}^B)^2$ and $(B_{Max}^C)^2$ could have
been $2.$ But this becomes impossible due to the restriction imposed by Eq.(\ref{bm3}). Moreover if any
 one set of tripartite marginals, say $P(a,b,d|x,y,w)$($W_B$ set) shows violation of the bilocal inequality,
the others set($W_C$) of marginals does not violate the bilocal inequality. So, generation of nonbilocal correlations
in one reduced network($\mathcal{N}_B$, say),
guarantees(considering generation of nonbilocal correlations up to detection by the sufficient criterion
provided by violation of the bilocal inequality(Eq.(\ref{A3l}))) absence of any such nonclassical feature(nonbilocality)
of quantum correlations in the other reduced network system($\mathcal{N}_C$). Hence, if maximal violation of
bilocal inequality is observed in one reduced network($\mathcal{N}_B$, say), then correlations from $\mathcal{N}_C$ cannot violate the bilocal
inequality(Eq.(\ref{A3l}))) and hence may not be nonbilocal. Such an observation is analogous to existing results
related to monogamy of quantum entanglement and nonlocality
(standard Bell-CHSH sense).
\section{Nonbilocal trade-off Relation}\label{trade}
Monogamy relation(Eq.(\ref{bm3})) guarantees existence of restrictions over distribution of nonbilocal correlations
 in reduced bilocal network systems. However, as already mentioned in the previous section, analogous to monogamy
 of nonlocal correlations in standard Bell-CHSH sense, measurement settings of nodal parties are kept fixed for
 assessing monogamy of nonbilocal correlations. However, in \cite{fei1}, it was pointed out that comparison of
  monogamy and trade-off relations of nonlocal correlations guarantees relaxation of restrictions over shareability of
  nonlocal correlations among bipartite reduced states. Such an observation is quite intuitive owing to the fact that
   in contrast to fixed measurement settings of the nodal party(considering nature of the bipartite marginals for
   sketching monogamy relation), for giving trade-off relation(connecting amount of Bell-CHSH violation by the bipartite
    marginals), the measurement settings for the nodal parties are not considered to be invariant. Hence optimization
 over parameters characterizing inputs of the nodal parties is possible separately while considering Bell-CHSH violation by each of the
  reduced states. For instance, it may so happen that Bell-CHSH violation by reduced state $\varrho_{AB}$(obtained from state $\varrho_{ABC}$),
  is optimized for one measurement direction($\vec{x_0}$, say) of Alice while the same by reduced state $\varrho_{AC}$ is optimized
   for some other measurement direction($\vec{x_1}\neq \vec{x_0}$, say) of Alice. \\
In this context, one may expect to encounter analogous observations in case of characterizing shareability of nonbilocal correlations.
However our findings guarantee somewhat counterintuitive feature.\\
\textit{Theorem.2}: Trade-off relation satisfied by upper bound of violation of bilocal inequality(Eq.(\ref{A3l})) by tripartite marginals in reduced networks $\mathcal{N}_B$ and $\mathcal{N}_C$ is same as the monogamy relation given by Eq.(\ref{bm3}).\\
\textit{Proof}: Each of Bob and Charlie performs separable measurements.  Measurement settings of the nodal parties Alice and Dick may vary while considering violation of bilocal inequality in each of the two reduced networks($\mathcal{N}_B$ and $\mathcal{N}_C$). Hence by Eq.(\ref{bm5}), we get:
$$(\mathbf{B}_{Max}^B)^2+(\mathbf{B}_{Max}^C)^2\leq $$
\begin{equation*}
\sqrt{\iota_1^B+\iota_2^B}\sqrt{\Lambda_1^B+\Lambda_2^B}+\sqrt{\iota_1^C+\iota_2^C}\sqrt{\Lambda_1^C+\Lambda_2^C}
\end{equation*}
For notations used in the proof, we refer to Appendix(Proof of theorem.1). Now applying A.M.$\geq$G.M. over the positive terms $\iota_1^B+\iota_2^B,$ $\iota_1^C+\iota_2^C,$ $\Lambda_1^B+\Lambda_2^B$ and $\Lambda_1^C+\Lambda_2^C,$ we get,
  $$ \frac{ \iota_1^B+\iota_2^B+\Lambda_1^B+\Lambda_2^B+
\iota_1^C+\iota_2^C+\Lambda_1^C+\Lambda_2^C}{2}.$$ Clearly the above expression is same as that given by Eq.(\ref{abm17}). Hence we get the same trade-off relation as that given by Eq.(\ref{bm3}).$\blacksquare$\\
The trade-off relation, being of the same form as that of the monogamy relation, restriction over shareability of nonclassical feature of quantum correlations(in terms of nonbilocality) is the same irrespective of whether measurement settings of the nodal parties remain fixed or not. Recent study on Bell-CHSH nonlocality reveals analogous findings regarding the fact that
 restrictions over distribution of nonclassical quantum correlations is independent of the fact whether nodal party's inputs are fixed or not. To be specific in \cite{hall1} a trade-off relation restriction shareability of nonlocal quantum correlations(Bell-CHSH)
  has been given which has the same form as that of a monogamy relation of Bell-CHSH nonlocality which was previously given in \cite{fei1}.\\
After ensuring existence of restriction over shareability of nonclassical correlations in quantum network scenario(characterized by source independence), we now discuss below practical significance of monogamy of nonbilocal correlations.
\section{Practical Implication}\label{prac}
Consider a network involving three parties Alice, Bob and Dick and two independent sources $\textbf{S}_1$ and $\textbf{S}_2.$ Monogamy of nonbilocality can be applied to design a secret bit sharing protocol involving the bilocal network secured against attacks of postquantum eavesdroppers. Below we give justification in support of our claim. \\
In \cite{cryp2}, Barrett \emph{etal.} proved a connection between the possibility of existence of a protocol secure against postquantum eavesdropping and quantum violation of Bell-CHSH inequality under nosignaling assumption. To be precise, they designed a protocol involving two parties(Alice and Bob, say). Alice and Bob share identical copies of entangled states. At the end of the protocol a secret bit is generated in between Alice and Bob although the source, generating the entangled states, is controlled by eavesdropper. Security of such a protocol is based on monogamy of Bell-CHSH violation by quantum correlations.  Now in our network scenario, as is clear from Eq.(\ref{abm14}), monogamy of nonbilocal correlations involves Bell violation by reduced states $\rho_{AB},$ $\rho_{AC},$
$\rho_{BD}$ and $\rho_{CD}.$ So restrictions over distribution of nonbilocal correlations involve restrictions over shareability of nonlocal correlations(in sense of Bell-CHSH violation) among the reduced states. This gives an indication about the possibility of designing a protocol, secured against eavesdroppers attack, via which secret bit can be generated in bilocal network scenario involving Alice, Bob(performing separable measurements) and Dick even if any eavesdropper(Charlie, say) who is capable of performing any separable measurement and have control over both the independent sources $\textbf{S}_1$ and $\textbf{S}_2.$
\section{Discussions}\label{conc}
Over years there has been thorough investigation of monogamous nature of quantum entanglement and quantum nonlocality in standard Bell-CHSH scenario. In this paper, we have considered bilocal quantum network scenario to investigate the same for some weaker form of quantum nonlocality(nonbilocality). Exploitation of our observations ensure monogamous nature of nonbilocal quantum correlations(up to existing detection criterion for nonbilocality(Eq.(\ref{A3l})))). Interestingly, restrictions over shareability of distribution of nonbilocal correlations among reduced networks($\mathcal{N}_B$ and $\mathcal{N}_C$) are the same irrespective of the inputs of the nodal parties(Alice and Dick) remaining fixed or not for observing violation of bilocal inequality in the reduced networks individually. From our discussions so far, it can be safely concluded that under the assumption of Bob and Charlie performing separable measurements, if quantum correlations in one reduced network($\mathcal{N}_B$, say) exhibit nonbilocality, then the correlations from the other one($\mathcal{N}_C$) cannot violate the bilocal inequality(Eq.(\ref{A3l})) and hence are bilocal(in terms of violating bilocal inequality). As we have already discussed, such monogamous nature of nonbilocal correlations can be utilized to design a secret sharing protocol secure against postquantum eavesdropper's attack. However  we have not been able to explicitly design any such protocol. One may find interest to develop one such protocol involving bilocal network. One may explore other means of applying this monogamous nature of quantum correlations in some other network related experimental tasks. Also study investigating inter-relation between monogamy of nonbilocality and other nonclassical aspects of quantum theory is a potential source of research activities.

\section{Appendix}
\begin{table}[htp]
\begin{center}
\begin{tabular}{|c|c|}
\hline
Party& Measurement setting\\
\hline
Alice& $\vec{\alpha_i}.\vec{\sigma}$\\
\hline
Bob& $\vec{\beta_i}^A.\vec{\sigma}\otimes\vec{\beta_i}^D.\vec{\sigma} $\\
\hline
 Charlie& $\vec{\gamma_i}^A.\vec{\sigma}\otimes\vec{\gamma_i}^D.\vec{\sigma} $\\
\hline
Dick& $\vec{\delta_i}.\vec{\sigma}$\\
\hline
\end{tabular}\\
\caption{The table gives the measurement settings of each of the four parties in the network. Two different values
of $i=0,1$ correspond to two arbitrary directions of projective measurements for each of the parties. As already
discussed in the main text, each of the two intermediate parties(Bob and Charlie) performs separable measurements\cite{raf1}. }
\end{center}
\label{table1}
\end{table}
\begin{widetext}
\textit{Proof of Theorem.1}: Both extreme and intermediate parties perform arbitrary projective
measurements(see Table.I). Applying Eq.(\ref{bm1}) to each of the two sets of marginals $W_B$ and $W_C,$ we get:\\
$$(\mathbf{B}^B)^2+(\mathbf{B}^C)^2=(\frac{1}{2}\small{\sum}_{i=0}^1\sqrt{|(\vec{\alpha_0}+
(-1)^i\vec{\alpha_1}).T_{AB}\vec{\beta_i^A}||\vec{\beta_i^D}.T_{BD}(\vec{\delta_0}+
(-1)^i\vec{\delta_1})|})^2$$
\begin{equation}\label{abm}
+(\frac{1}{2}\small{\sum}_{i=0}^1\sqrt{|(\vec{\alpha_0}+
(-1)^i\vec{\alpha_1}).T_{AC}\vec{\gamma_i^A}||\vec{\gamma_i^D}.T_{CD}(\vec{\delta_0}+
(-1)^i\vec{\delta_1}|}))^2,
\end{equation}
with $T_{AB},$ $T_{AC},$ $T_{BD}$ and $T_{CD}$ denoting the correlation tensor(Eq.(\ref{st4i})) of reduced bipartite
 states $\textmd{Tr}_C(\rho_{ABC}),$ $\textmd{Tr}_B(\rho_{ABC}),$ $\textmd{Tr}_C(\rho_{BCD})$ and
 $\textmd{Tr}_B(\rho_{BCD})$ respectively($\textmd{Tr}_C(\rho_{ABC})$ stands for the reduced state obtained
 after tracing out party $C$ from tripartite state $\rho_{ABC}$).
Following the procedure used in \cite{hord1}, we introduce two pairs(one for Alice and other for Dick's
measurement settings) of mutually orthogonal unit vectors:
\begin{equation}\label{abm1}
   \vec{\alpha_0}+(-1)^i\vec{\alpha_1}=2\cos(i\small{\frac{\pi}{2}}-\mu)\vec{\upsilon^A_i},\,\,i\in\{0,1\},
   \,\,\mu\in[0,\pi]
\end{equation}
and
\begin{equation}\label{abm2}
     \vec{\delta_0}+(-1)^i\vec{\delta_1}=2\cos(i\small{\frac{\pi}{2}}-\nu)\vec{\upsilon^D_i},\,\,i\in\{0,1\},
     \,\,\nu\in[0,\pi].
\end{equation}
By Eqs.(\ref{abm1},\ref{abm2}), we get from Eq.(\ref{abm}),
$$(\mathbf{B}^B)^2+(\mathbf{B}^C)^2=(\small{\sum}_{i=0}^1\sqrt{|(\vec{\upsilon_i^A}.T_{AB}\vec{\beta_i^A})
(\vec{\beta_i^D}.T_{BD}\vec{\upsilon_i^D})|
|\cos(i\small{\frac{\pi}{2}}-\mu)\cos(i\small{\frac{\pi}{2}}-\nu)|})^2$$
\begin{equation*}
+(\small{\sum}_{i=0}^1\sqrt{|(\vec{\upsilon_i^A}.T_{AC}\vec{\gamma_i^A})(\vec{\gamma_i^D}.T_{CD}
\vec{\upsilon_i^D})|
|\cos(i\small{\frac{\pi}{2}}-\mu)\cos(i\small{\frac{\pi}{2}}-\nu)|})^2.
\end{equation*}
$$\,\,\,\,\,\,\,=(\small{\sum}_{i=0}^1\sqrt{|(\vec{\beta_i^A}.T^T_{AB}\vec{\upsilon_i^A})(\vec{\beta_i^D}.
T_{BD}\vec{\upsilon_i^D})|
|\cos(i\small{\frac{\pi}{2}}-\mu)\cos(i\small{\frac{\pi}{2}}-\nu)|})^2$$
\begin{equation}\label{abm3}
+(\small{\sum}_{i=0}^1\sqrt{|(\vec{\gamma_i^A}.T^T_{AC}\vec{\upsilon_i^A})(\vec{\gamma_i^D}.T_{CD}
\vec{\upsilon_i^D})|
|\cos(i\small{\frac{\pi}{2}}-\mu)\cos(i\small{\frac{\pi}{2}}-\nu)|})^2.
\end{equation}
Now each of $\vec{\beta_i^A},$ $\vec{\beta_i^D},$ $\vec{\gamma_i^A}$ and $\vec{\gamma_i^D}(i=0,1)$
being unit vectors, applying Cauchy-Schwarz's inequality, we get:
$$(\mathbf{B}^B)^2+(\mathbf{B}^C)^2\leq(\small{\sum}_{i=0}^1\sqrt{||T^T_{AB}\vec{\upsilon_i^A}|||
|T_{BD}\vec{\upsilon_i^D}||
|\cos(i\small{\frac{\pi}{2}}-\mu)\cos(i\small{\frac{\pi}{2}}-\nu)|})^2$$
\begin{equation}\label{abm4}
+(\small{\sum}_{i=0}^1\sqrt{||T^T_{AC}\vec{\upsilon_i^A}||||T_{CD}\vec{\upsilon_i^D}||
|\cos(i\small{\frac{\pi}{2}}-\mu)\cos(i\small{\frac{\pi}{2}}-\nu)|})^2.
\end{equation}
Let $Q_i^B$$=$$||T^T_{AB}\vec{\upsilon_i^A}||||T_{BD}\vec{\upsilon_i^D}||$ and $Q_i^C$$=$$||T^T_{AC}
\vec{\upsilon_i^A}||||T_{CD}\vec{\upsilon_i^D}||(i=0,1).$ Then Eq.(\ref{abm4}) becomes,
\begin{equation}\label{abm5}
 (\mathbf{B}^B)^2+(\mathbf{B}^C)^2\leq(\small{\sum}_{i=0}^1\sqrt{Q_i^B
|\cos(i\small{\frac{\pi}{2}}-\mu)\cos(i\small{\frac{\pi}{2}}-\nu)|})^2+(\small{\sum}_{i=0}^1\sqrt{Q_i^C
|\cos(i\small{\frac{\pi}{2}}-\mu)\cos(i\small{\frac{\pi}{2}}-\nu)|})^2
\end{equation}
Expanding right hand side of the inequality(Eq.(\ref{abm5})), we get:
$$(\mathbf{B}^B)^2+(\mathbf{B}^C)^2\leq (Q_0^B+Q_0^C)|\cos(\mu)\cos(\nu)|+(Q_1^B+Q_1^C)|\sin(\mu)\sin(\nu)|$$
\begin{equation}\label{abm6}
+2(\sqrt{Q_0^B*Q_1^B}
+\sqrt{Q_0^C*Q_1^C})\sqrt{|\cos(\mu)\cos(\nu)\sin(\mu)\sin(\nu)|}
\end{equation}
It is maximized for $\mu=n\pi\pm\nu.$ Hence Eq.(\ref{abm6}) gives:
$$(\mathbf{B}^B)^2+(\mathbf{B}^C)^2\leq (Q_0^B+Q_0^C)\cos^2(\nu)+(Q_1^B+Q_1^C)\sin^2(\nu)$$
\begin{equation*}
+2(\sqrt{Q_0^B*Q_1^B}
+\sqrt{Q_0^C*Q_1^C})|\cos(\nu)\sin(\nu)|
\end{equation*}
\begin{equation*}
=(\sqrt{Q_0^B}|\cos(\nu)|+\sqrt{Q_1^B}|\sin(\nu)|)^2+(\sqrt{Q_0^C}|\cos(\nu)|+\sqrt{Q_1^C}|\sin(\nu)|)^2
\end{equation*}
\begin{equation}\label{abm7}
=A_B(\nu)+A_C(\nu)
\end{equation}
where $A_B(\nu)$$=$$(\sqrt{Q_0^B}|\cos(\nu)|+\sqrt{Q_1^B}|\sin(\nu)|)^2$ and $A_C(\nu)$$=$$(\sqrt{Q_0^C}|\cos(\nu)|+\sqrt{Q_1^C}|\sin(\nu)|)^2$ are connected with network $\mathcal{N}_B$ and $\mathcal{N}_C$ respectively. Now without of any loss of generality, consider the term $A_B(\nu).$ \\
\begin{equation*}
A_B(\nu)=(\sqrt{Q_0^B}|\cos(\nu)|+\sqrt{Q_1^B}|\sin(\nu)|)^2
\end{equation*}
\begin{equation*}
    \leq (\sqrt{(\sqrt{Q_0^B})^2+(\sqrt{Q_1^B})^2})^2
\end{equation*}
\begin{equation}\label{abm8}
  =  Q_0^B+Q_1^B
\end{equation}
The above inequality is obtained by applying the inequality $x\cos(\theta)+y\sin(\theta)\leq \sqrt{x^2+y^2}.$
Let $A_B(\nu)$ be maximized for $\nu=\nu_1.$ As already argued, for deriving monogamy relations, measurement settings of Alice and Dick should be same in both the networks $N_B$ and $N_C.$ Now for $\nu=\nu_1,$ \\
\begin{equation*}
A_C(\nu_1)=(\sqrt{Q_0^C}|\cos(\nu_1)|+\sqrt{Q_1^C}|\sin(\nu_1)|)^2
\end{equation*}
\begin{equation*}
    \leq (\sqrt{(\sqrt{Q_0^C})^2+(\sqrt{Q_1^C})^2})^2
\end{equation*}
\begin{equation}\label{abm9}
   = Q_0^C+Q_1^C
\end{equation}
Equality holds if $A_C(\nu)$ is maximum for $\nu=\nu_1.$ Hence, by Eqs(\ref{abm8},\ref{abm9}), we get,
\begin{equation}\label{abm10}
(\mathbf{B}^B)^2+(\mathbf{B}^C)^2\leq Q_0^B+Q_0^C+Q_1^B+Q_1^C   
\end{equation}
By our argument, till now, we have shown that if anyone of $A_B(\nu)$ or $A_C(\nu)$ attains maximum for some value of $\nu,$ then $A_B(\nu)+A_C(\nu)$$\leq$$G$ where $G$$=$$Q_0^B+Q_0^C+Q_1^B+Q_1^C.$ If possible, there exists some other fixed value of $\nu$, say $\nu_2$ such that,
\begin{equation}\label{abm11}
    A_B(\nu_2)+A_C(\nu_2)>G
\end{equation}
 But,
\begin{equation*}
A_B(\nu_2)=(\sqrt{Q_0^B}|\cos(\nu_2)|+\sqrt{Q_1^B}|\sin(\nu_2)|)^2
\end{equation*}
\begin{equation*}
   \leq  Q_0^B+Q_1^B
\end{equation*}
Equality holds if $\nu_1=\nu_2.$ Similarly, $A_C(\nu_2)$$\leq$$  Q_0^C+Q_1^C.$ These relations in turn give,
\begin{equation}\label{abm12}
    A_B(\nu_2)+A_C(\nu_2)\leq G
\end{equation}
which contradicts Eq.(\ref{abm11}). Hence when maximized over all possible values of measurement parameter $\nu,$ we get, Eq.(\ref{abm10}). \\
Put $R_i^B=||T_{BD}\vec{\upsilon_i^D}||$ and $R_i^C=||T_{CD}\vec{\upsilon_i^D}||(i=0,1).$ Inequality given by Eq.(\ref{abm10})
becomes,
\begin{equation}\label{abm13}
(\mathbf{B}^B)^2+(\mathbf{B}^C)^2\leq \sum_{i=0}^1(R_i^B||T^T_{AB}\vec{\upsilon_i^A}||+R_i^C||T^T_{AC}\vec{\upsilon_i^A}||)
\end{equation}
Now, it was shown in \cite{raf1}, that for any matrix $U$ and for any vector $\vec{v}, $$||U\vec{v}||^2=\vec{v}.
U^TU\vec{v}$. Again for any matrix $U,$ $U^TU$ is always diagonlizable. Hence both $T^T_{AB}T_{AB}$ and $T^T_{AC}T_{AC}$ are both diagonlizable. Let $\Lambda_1^B\geq\Lambda_2^B\geq \Lambda_3^B$ be the eigen values of $T^T_{AB}T_{AB}.$ Analogously let $\Lambda_1^C\geq\Lambda_2^C\geq \Lambda_3^C$ be the eigen values of $T^T_{AC}T_{AC}$ respectively.
For the term $||T^T_{AB}\vec{\upsilon_i^A}||,$  expressing $\vec{\upsilon_i^A}$ in the eigen vector basis of
$T^T_{AB}T_{AB}$ and likewise expressing $\vec{\upsilon_i^A}$ in the eigen vector basis of $T^T_{AC}T_{AC}$ in
 the term $||T^T_{AC}\vec{\upsilon_i^A}||(i=0,1),$
\begin{equation}\label{abm14}
(\mathbf{B}^B)^2+(\mathbf{B}^C)^2\leq \sum_{j=0}^1(R_j^B\sqrt{\small{\sum}_{i=0}^2\Lambda_{i+1}^B(\upsilon_{j,i}^{AB})^2}+
    R_j^C\sqrt{\small{\sum}_{i=0}^2\Lambda_{i+1}^C(\upsilon_{j,i}^{AC})^2})
\end{equation}
where $\vec{\upsilon_i^A}$$=$$\vec{\upsilon_i^{AC}}$(representation in eigen vector basis of $T^T_{AC}T_{AC}$) and $\vec{\upsilon_i^A}$$=$$\vec{\upsilon_i^{AB}}$(representation in eigen vector basis of $T^T_{AB}T_{AB}$).
Now due to the orthogonality constraint over the unit vectors $\vec{\upsilon_0^{AB}},$ $\vec{\upsilon_1^{AB}}$ and similarly that of $\vec{\upsilon_0^{AC}},$ $\vec{\upsilon_1^{AC}}$  and also
due to $\Lambda_1^B\geq\Lambda_2^B\geq \Lambda_3^B$ and $\Lambda_1^C\geq\Lambda_2^C\geq \Lambda_3^C,$ maximization
over Alice's measurement settings(maximum being obtained for $\vec{\upsilon_0^{AB}}=\vec{\upsilon_0^{AC}}=(1,0,0)$ and $\vec{\upsilon_1^{AB}}=\vec{\upsilon_1^{AC}}=(0,1,0))$, give:
\begin{equation}\label{abm15}
(\mathbf{B}^B)^2+(\mathbf{B}^C)^2\leq \sum_{j=0}^1(R_j^B\sqrt{\Lambda_{j+1}^B}+
   R_j^C\sqrt{\Lambda_{j+1}^C})
\end{equation}
Now maximizing over Dick's measurement settings in an analogous approach, we get
\begin{equation}\label{abm16}
 (\mathbf{B}^B_{Max})^2+(\mathbf{B}^C_{Max})^2= \sum_{j=1}^2(\sqrt{\iota_j^B}\sqrt{\Lambda_j^B}+\sqrt{\iota_j^C}\sqrt{\Lambda_j^C})
\end{equation}
where $\iota_1^B\geq\iota_2^B\geq \iota_3^B$ and $\iota_1^C\geq\iota_2^C\geq \iota_3^C$ are the eigen values of
$T^T_{BD}T_{BD}$ and $T^T_{CD}T_{CD}$ respectively. Now, by applying, A.M.$\geq$G.M. for positive integers, $\iota_i^B,\iota_i^C,$ $\Lambda_i^B$ and $\Lambda_i^C(i=1,2),$ we get:
\begin{equation}\label{abm17}
(\mathbf{B}^B_{Max})^2+(\mathbf{B}^C_{Max})^2= \frac{ \iota_1^B+\iota_2^B+\Lambda_1^B+\Lambda_2^B+
\iota_1^C+\iota_2^C+\Lambda_1^C+\Lambda_2^C}{2}.
\end{equation}
Now for the state $\rho_{ABC}$(generated by the source $\textbf{S}_1$) which is shared between Alice, Bob
and Charlie\cite{hall1},
\begin{equation}\label{abm18}
\Lambda_1^B+\Lambda_2^B+\Lambda_1^C+\Lambda_2^C\leq 2,
\end{equation}
Analogously, for the state $\rho_{BCD}$(generated by the source $\textbf{S}_2$), we get,
\begin{equation}\label{abm19}
\iota_1^B+\iota_2^B+\iota_1^C+\iota_2^C\leq 2,
\end{equation}
Using, Eqs.(\ref{abm18},\ref{abm19}), in Eq.(\ref{abm17}), we get the required monogamy relation(Eq.(\ref{bm3})).
\end{widetext}
\end{document}